\documentclass[10pt,letterpaper,twocolumn]{article}	% two col.

\usepackage{graphicx}

\usepackage{epstopdf}

\def\lesssim{\mathrel{\hbox{\rlap{\hbox{\lower4pt\hbox{$\sim$}}}\hbox{$<$}}}}
\def\gtrsim{\mathrel{\hbox{\rlap{\hbox{\lower4pt\hbox{$\sim$}}}\hbox{$>$}}}}

\begin{document}
\title{Three Disk
Oscillation Modes of 
Rotating Magnetized Neutron Stars}

\author{R.V.E. Lovelace \\
Departments of Astronomy and  Applied and
Engineering Physics, \\
Cornell University, Ithaca, NY 14853-6801; 
RVL1@cornell.edu 
\and
M.M. Romanova \\
Department of Astronomy,\\
Cornell University, Ithaca, NY 14853-6801;
romanova@astro.cornell.edu}

\maketitle

\begin{abstract}

We discuss three  specific modes of accretion disks 
around rotating magnetized neutron stars which may
explain the separations of the kilo Hertz
quasi periodic oscillations (QPO) seen in
low mass X-ray binaries.   
  The existence of  these modes requires 
that there be a maximum
in the angular velocity of the accreting material,
and that the fluid is in
stable, nearly circular motion near this
maximum rather than moving
rapidly towards the star or out of the disk plane
into funnel flows.   
   It is presently not known if these
conditions occur, but  we are exploring 
 this with 3D magnetohydrodynamic
simulations and will report the results elsewhere. 
  The first mode is a corotation mode which is
radially trapped in the vicinity of the maximum of the 
disk rotation rate and is unstable.
   The second mode, relevant to relatively
slowly rotating stars, is a magnetically driven
eccentric ($m=1$)  oscillation of the 
disk excited at a Lindblad radius in the vicinity
of the maximum of the disk  rotation.
   The third mode, relevant to rapidly rotating
stars, is a magnetically coupled
eccentric ($m=1$) and an axisymmetric ($m=0$)
radial disk perturbation which has an
inner Lindblad radius also in the vicinity
of the maximum of the disk rotation.
   We suggest that the first mode is
associated with the upper QPO frequency, $\nu_u$,  the
second with the lower QPO frequency, $\nu_\ell =\nu_u-\nu_*$,
and  the third with the lower QPO frequency, 
$\nu_\ell=\nu_u-\nu_*/2$, where $\nu_*$ 
is the star's rotation rate.

\end{abstract}

\noindent{keywords:  accretion, accretion disks ---  stars: neutron
--- X-rays: binaries --- magnetohydrodynamics}

\section{Introduction}

 Low mass X-ray binaries often display twin
kilo-Hertz quasi-periodic oscillations (QPOs)
in their X-ray emissions (van der Klis 2006; Zhang
et al. 2006).
   A wide variety of different models have
been proposed to explain the origin and correlations
of the different QPOs.  
    These include
the beat frequency model (Miller, Lamb, \& Psaltis 1998;
Lamb \& Miller 2001; Lamb \& Miller 2003), the
relativistic precession model (Stella \& Vietri 1999),
the Alfv\'en wave model (Zhang 2004), and warped disk
models (Shirakawa \& Lai 2002; Kato 2004).

   A puzzling aspect of the some of the twin QPO
sources considered in this work is that
the  difference between the upper
$\nu_u$ and lower $\nu_\ell$ QPO frequencies is
roughly either the spin frequency of the star $\nu_*$
($3$ cases where $\nu_*=270,~330,~\&~ 363$ Hz) 
or one-half this frequency,
$\nu_*/2$ ($4$ cases where 
$\nu_*=401,~524,~581,~\&~619$ Hz), for the
cases where
$\nu_*$ is known, even though
$\nu_u$ and $\nu_\ell$ vary significantly
(see, e.g., Zhang et al. 2006). 
  A further type of behavior appears in
the source Cir X-1 (Boutloukos et al. 2006), but
this is not considered here.
  The cases where $\nu_u-\nu_\ell \approx \nu_*$
may be explained by the beat frequency
model (Miller et al. 1998), but the explanation
of the cases where $\nu_u-\nu_\ell \approx \nu_*/2$
is obscure.

  Section 2.1 discusses the corotation instability,  
\S 2.2  the exccentric ($m=1$)
mode of the disk driven by the star's rotating
magnetic field, and \S 2.3 the coupled exccentric
plus axisymmetric mode ($m=0~\&~1$) also due to the star's
rotating magnetic field.
Section 4 gives the conclusions.

\section{Three Modes}

\subsection{Corotation Instability}

    We assume a  pseudo-Newtonian  
gravitational potential $\Phi_g=-GM_*/(r-r_S)$,
where $M_*$ is the star's mass and $r_S\equiv 2GM_*/c^2$.
In the absence of the star the angular
velocity of disk matter  is $\Omega_{g\phi}=
\{GM_*/[r(r-r_S)^2]\}^{1/2}$ for $r\geq 3r_S$. 
   Near the star  the disk's
angular rotation rate in the 
equatorial plane is modeled as
\begin{equation}
\Omega_\phi(r) ={\Omega_* f(r) \over 1+f(r)}
+ {\Omega_{g\phi}(r) \over 1+f(r)}~,
\end{equation}
where $f(r)=\exp[-(r-r_0)/\Delta]$ with $r_0$
the standoff distance of the boundary layer
and $\Delta$  its thickness which are expected
to depend on the accretion rate and  the
star's magnetic field. 
  The radial force equilibrium in the midplane of
an axisymmetric
disk is $\rho r(\Omega_\phi^2-\Omega_{g\phi}^2) 
= d(p+B^2/8\pi)/dr$, where $\rho$ is the midplane
density and midplane field ${\bf B}=B(r)\hat{\bf z}$.  
   Figure 1 shows the equilibrium quantities for
an illustrative case.
   Clearly, $\Omega_\phi(r)$ has a maximum value
outside of the star at a distance denoted $r_m \sim r_0$. 
   The importance of this maximum for models of
QPOs  was discussed
earlier by Alpar and Psaltis (2005). 
   Three dimensional magnetohydrodynamic (MHD) simulations
of disk accretion to rotating  magnetized stars  can
in principle be used to determine
$\Omega_\phi(r)$ for different conditions (Romanova,
Kulkarni, \& Lovelace 2007).
    However, for the present purposes 
the dependence of equation (1)
is used.  
  It is needed only 
for distances $r\gtrsim r_m$ as discussed
below. 
    In this region the radial epicyclic
frequency is  $\Omega_r(r)=[r^{-3}d(r^4\Omega_\phi^2)/dr]^{1/2}$.
   The maximum value of $\Omega_\phi(r)$ has
the approximate dependence max$(\Omega_\phi/2\pi)
\approx 2040(3r_S/r_0)^{1.74}$ Hz for $ 3 < r_0/r_S <5$,
$\Delta/r_S =0.05$, and $M_*=1.4M_\odot$. 
   For this range of $r_0$, max($\Omega_\phi$)
changes by a factor of $2.4$.

 %%%%%%%%%%%%%%%%%%%%%%%%%%%%%%%%%%%%%%%%%%
\begin{figure}[hbtp]
\centerline{\includegraphics[scale=0.5]{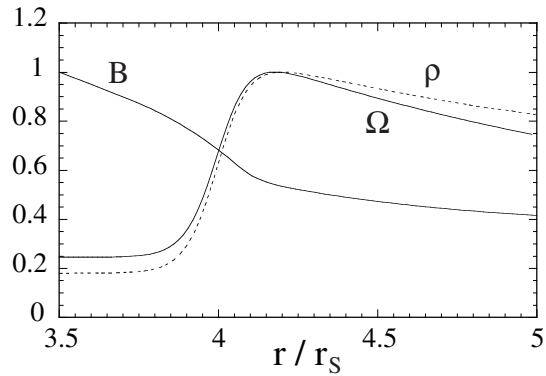}}
\caption{Equilibrium profiles of the main
variables normalized to their maximum values.
 For this case, $M_*=1.4M_\odot$,
$\nu_*=\Omega_*/2\pi=300$~Hz,
$r_S \approx 4.14\times10^5$ cm, $r_0=4 r_S$, $\Delta=0.05r_S$,
$c_s=0.1 r \Omega_\phi$, $\rho_{\rm max} \approx 0.024$ g/cm$^3$,
$B(r/r_S=3.5)\approx 1.7\times 10^9$ G, and ${\rm max}(\Omega_\phi/2\pi)
\approx 1220$ Hz. }
\end{figure}
%%%%%%%%%%%%%%%%%%%%%%%%%%%%%%%%%%%%%%%%%%

  We consider a WKB treatment of the
corotation ($\omega \approx m\Omega_\phi$) 
or Rossby type wave of
the disk with pressure perturbation
$$
\delta p\sim \exp\bigg(i\int^r dr
k+im\phi-i\omega t\bigg)~,
$$
where   $k$ is the radial wavenumber,
$m=1,2,..$, and $\omega=\omega_r+i\omega_i$, with $\omega_r$
the angular frequency of the perturbation and $\omega_i$ the 
growth rate. 
   For the conditions of Figure 1, the Alfv\'en speed
$c_A=B/(4\pi\rho)^{1/2}$ is much larger than the sound
speed $c_s$.  Also, we assume and verify later that
$|\omega-m\Omega_\phi|^2 \ll \Omega_r^2$.
    Under these conditions
    \vfill\eject
$$
k^2(r)= -\left({m\over r}\right)^2
- \left({\Omega_r \over c_A}\right)^2
-{1\over L_B^2} 
-{\Omega_\phi \over r {\cal F}}{d \over dr }
\left( {r{\cal F} \over \Omega_\phi L_B}\right)
$$
\begin{equation}
-{2 m \over r}\left({2\over L_B}\! 
+\!{1\over {\cal F}}{d{\cal F}\over dr}\right)
\Re\left({\Omega_\phi \over \Delta \omega}\right)
-\left({m \over r}\right)^2{ c_A^2 \over L_B L_b}
\Re\left({1\over \Delta \omega^2}\right),
\end{equation}
where $\Delta\omega \equiv \omega-m\Omega_\phi$,
  $L_B^{-1}\equiv d\ln(B/\rho)/dr$, 
$L_b^{-1} \equiv d \ln(B)/dr$, 
${\cal F} \equiv \rho \Omega_\phi/\Omega_r^2$,
which all depend on $r$, and
$\Re(..)$ denotes the real part (Lovelace, Turner, \& Romanova 2007).
Equation (2) generalizes the calculation of
Lovelace et al. (1999) to include
the influence of the magnetic field perturbation
$\delta {\bf B}=\delta B(r,\phi,t)\hat{\bf z}$.

   Figure 2 shows the radial dependence of $k^2$
for a representative case.
   For the chosen value $\omega_r/2\pi =1100$~Hz,
which is somewhat less than the maximum value of
$\Omega_\phi(r)$, and $m=1$,
it is seen that $k^2(r)\geq 0$ in a finite
radial interval in the vicinity of the maximum
of $\Omega_\phi(r)$.  
   Thus the wave is radially trapped in the vicinity of
the maximum of $\Omega_\phi(r)$.
   Analogous radially trapped modes were analyzed
earlier by Lovelace et al. (1999) and Li et al. (2000)
and verified in two dimensional hydrodynamic
simulations (Li et al. 2001).
  The Bohr-Sommerfeld quantization 
condition $\int_{r({\rm in})}^{r({\rm out})} dr~\! k =
(n+1/2)\pi$, $n=0,1,..$ allows the determination
of the growth rate $\omega_i$.
For the case shown, $\omega_i/2\pi =\nu_i \approx 55$~Hz
for $n=0$ which gives the  largest growth rate,
and $r({\rm in})/r_S=4.08$, $r({\rm out})/r_S =4.48$. 
   The growth rate increases as $\Delta$ decreases.
Similar values of the growth rates are  found for
$\omega_r$ somewhat less than  $m~\!{\rm max}(\Omega_\phi)$ for
$m=2,3,..$.  
   The nonlinear saturation of the growth of the
modes can in principle be found by MHD simulations
(Koldoba et al. 2002; Romanova et al. 2007).

    From Figure 2 we see that the validity of equation (1) is
needed only from the vicinity of the maximum of $\Omega_\phi(r)$,
that is, from $r/r_S = 4.08$ where
$\Omega_\phi/{\rm max}(\Omega_\phi)=0.94$ 
(the inner turning point) and outward (including
the outer turning point at $r/r_S=4.48$).

   Owing to the perturbation, the surface temperature of
the disk is $T(r,\phi,t)=$
\begin{equation}
T_0+\delta T_1\exp(i\phi-\omega_1 t)
+\delta T_2(r)\exp(2i\phi-i\omega_2 t)+~.~.~,
\end{equation}
where $T_0(r)$ is the unperturbed temperature,
$\delta T_{1,2}(r)\ll T_0$ are the amplitudes of
the $m=1,2$ corotation modes, and $\omega_{1,2}$
are their frequencies.
The corresponding flux density is proportional to
$S(r,\phi,t)\sim T_0^4+4T_0^3\delta T_1\exp(i\phi-\omega_1t)+..$
The total  flux for a face-on disk, $L\sim \int rdrd\phi S$,
is independent of time.  For a more general disk
orientation, the Doppler effect due to the
disk rotation gives a boost for say $\phi=0$  and
a decrement for $\phi=\pi$.  
   This corresponds to multiplying $S$ by
$[1+ \epsilon \exp(-i\phi)]$,  
with $\epsilon(r) \ll 1$.
    Consequently, there is
a contribution to the source
flux $\delta  L \sim \int rdr d\phi
\exp(-i\phi) S(r,\phi,t) \sim  
4 \int rdr T_0^3\epsilon \delta
T_1(r)\exp(-i\omega_1t)$.  
   As explained in the next section,
we interpret this frequency  as the
upper frequency component of the twin QPOs.

 %%%%%%%%%%%%%%%%%%%%%%%%%%%%%%%%%%%%%%%%%%
\begin{figure}[hbtp]
\centerline{\includegraphics[scale=0.5]{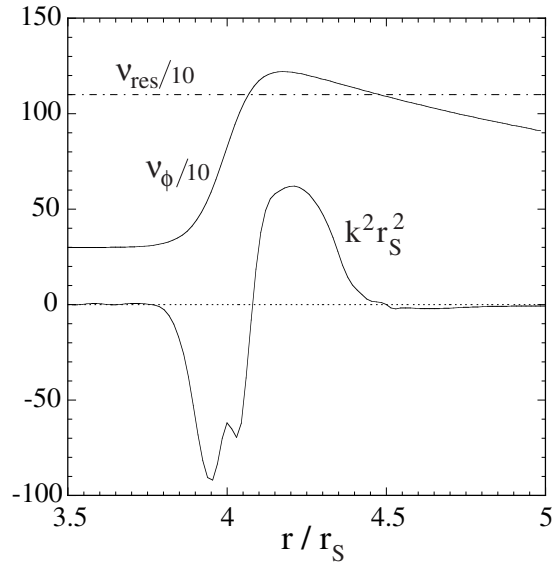}}
\caption{Radial dependences of the
different frequencies $\nu_\phi =\Omega_\phi/2\pi$,
the resonant frequency
$\nu_{res}=\omega_r/2\pi=1100$~Hz, and the square of
the radial wavenumber, $k^2$ obtained from
equation (2) for $m=1$ and for the same conditions as for Figure 1.}
\end{figure}
%%%%%%%%%%%%%%%%%%%%%%%%%%%%%%%%%%%%%%%%%%

\subsection{Magnetically Driven $m=1$ Mode}

   We now give a heuristic treatment of
perturbations of the disk
excited by the star's rotating, non-axisymmetric magnetic
field.  
   One component of a general perturbation
is described by the radial displacement
of the disk matter,  ${\cal E}(r,\phi, t)$ (e.g., Zhang \&
Lovelace 2005)
     The
equation for ${\cal E}$ is simply
\begin{equation}
{ d^2 {\cal E} \over dt^2} = -\big(\Omega_r^2+k^2c_s^2\big)
{\cal E} +
\delta F(r,\phi,t)~,
\end{equation}
where
$
{d / dt} \equiv  {\partial / \partial t} + \Omega_\phi
{\partial/ \partial \phi},$
and $\delta F$ is the radial force due to the star's magnetic
field.
  The radial oscillation
frequency (squared) on the right-hand side of
the equation consists
$\Omega_{r}^2$ as given earlier plus a thermal contribution
$ k^2 c_s^2,$
where $k$ is the radial wavenumber of the perturbation
in the WKB approximation.

    We consider the case where the star has a
small size or point-like dipole located close to the star's
surface near the rotation axis with its magnetic moment
parallel to the star's surface. 
  Ruderman (2006) gives  physical arguments for this type of
configuration of spun-up neutron stars.
   At a given radial distance in the
equatorial plane, the total magnetic field
$ B_z= B_z^v + B_z^i$ consists of  the
vacuum component
  $B_z^v =B_0(r) \cos(\phi - \Omega_* t)$,
with $\Omega_*$ the star's angular rotation rate, 
and the induced component, $B_z^i$.  
  The component $B_z^v$ acts to drive a current
flow in the disk ${\cal K}_\phi$ which in turn gives rise to the
induced magnetic field of the disk $B_z^i$.
    The radial force  includes a coherent
component $\propto B_z^v$ which can be written as 
$
\delta F =C_1\exp(i\phi -i\Omega_* t),
$ 
where $C_1(r)=\langle{\cal K}_\phi\rangle B_0/(c\Sigma)$
with the average is over $\phi$. 
   Considering only the coherent component,
equation (4) is then a driven oscillator.
   With $ik \rightarrow d/dr$ and 
${\cal E} \sim \exp(i\phi-i\Omega_* t)$,
it becomes
\begin{equation}
\left({d^2 \over dr^2} -{\cal D}(r)\right){\cal E}= -C_1/c_s^2~,
\end{equation}
with ${\cal D}\equiv [\Omega_r^2-(\Omega_\phi-\Omega_*)^2]/c_s^2$.
  Depending mainly on the value of $\Omega_*$,
there may be
Lindblad resonances with inner and
outer Lindblad radii $r_{Li,o}$ where
${\cal D}(r_{Li,o})=0$.  
  The region between the two radii is permitted
in the sense that ${\cal D}\leq 0$ while the regions
$r>r_{Lo}$ and $r<r_{Li}$ are forbidden.
  The existence of the Lindblad resonances means that a
weak magnetic disturbance  can give rise to a strong
disk response  in the vicinity of $r_{L}$
proportional to $C_1(r_L)$ (Goldreich
\& Tremaine 1979).
  Because of the rapid decrease of the magnetic
field the response at the inner Lindblad resonance
is expected to be stronger than that at the outer
resonance.  
   However, the solution of equation (5) is beyond the scope
of this work.
   We find that there
are Lindblad resonances only for
for $\nu_*=\Omega_*/2\pi < 380$~Hz for the same conditions
as Figure 2.  
   For higher $\nu_*$, all values of $r$ are
forbidden.
   For $\nu_*=300$~Hz, 
 $r_{Li}/r_S\approx 4.34$, 
which is within the region of the mentioned corotation
instability, and $r_{Lo}/r_S \approx 4.83$
which is outside the region of corotation
instability.
  The disk velocity
$\Omega_\phi r$ is supersonic relative to 
the velocity of the perturbation $\Omega_* r$ for  
 $r$ larger than $r_{Li}$.  
Consequently the perturbation is
a leading spiral wave with $k <0$.  
   At the outer Lindblad
radius the disk velocity is also supersonic
relative to $\Omega_* r$, and excitation of the disk
at this radius gives a trailing spiral,
$k>0$ for $r < r_{Lo}$.

    The  interaction
between the corotation perturbation and the magnetic
perturbation of the disk is in general
nonlinear.   The perturbed disk surface
temperature can be represented as a product
of the two perturbations
$\tilde T(r,\phi,t) \sim [1+\epsilon_{M}
|{\cal E}/r|\exp(-i\phi+i\Omega_* t)][T_0+\delta
T_1\exp(i\phi-i\omega_1 t)+..]=T_0+\epsilon_M 
|{\cal E}/r|\delta T_1
\exp[-i(\omega_1-\Omega_*)t]+..$, where 
$\epsilon_M \leq  1$. 
       Consequently, there is
a contribution to the source
flux $\delta  L \sim \int rdr d\phi
\exp(-i\phi) S(r,\phi,t) \sim  
4 \int rdr T_0^3\epsilon_M |{\cal E}/r| \delta
T_1(r)\exp[-i(\omega_1-\Omega_*)t]$, where
$S\sim \tilde T^4$.  
   For $\nu_* < \sim 380$~Hz, we interpret 
$\omega_1-\Omega_*$ as the
lower frequency component of twin QPOs.

\subsection{Magnetically Coupled $m=0~\&~1$ Mode}

  Consider now higher rotation frequencies $\nu_*$.
Note  that the radial force perturbation  includes a
contribution of the form
$\delta F \sim \Sigma^{-1}\Re( {\cal E})\partial 
({\cal K}_\phi
B_z/c)/\partial r$ which  again has a coherent term
proportional to $B_z^v$.
    We consider only the coherent term which can
be written as
$\delta F =\Re( {\cal E}) D_1 \cos(\phi - \Omega_*t)$,
where $D_1= \partial \langle{\cal K}_\phi B_0\rangle
/\partial r/(c\Sigma)=$ real.
In this case ${\cal E}$ necessarily consists of
different $\phi-$harmonics.  That is,
 ${\cal E}={\cal E}_0 \exp(-i\omega t) +{\cal E}_1
\exp(i\phi-i\omega t)+{\cal E}_2\exp(2i\phi-\omega t)+..$.
We find
\begin{eqnarray}
\left(c_s^2{d^2 \over
dr^2}-\big[\Omega_r^2-\omega^2\big] \right){\cal E}_0 
&=&-{1\over 2} {\cal E}_1^*D_1~,
\nonumber\\
\left(c_s^2{d^2 \over
dr^2}-\big[\Omega_r^2-(\Omega_\phi-\omega)^2 \big]
\right){\cal E}_1 &=& -{1\over 2}{\cal E}_0^* D_1~,
\end{eqnarray}
where necessarily $\omega =\Omega_*/2$.
  For the radii of interest in the vicinity of $r_0$, 
$\Omega_r^2 \gg \omega^2 $, so that the right-hand
side of the second equation is 
$\approx -D_1^2{\cal E}_1/(4\Omega_r^2)$.
   We identify $|D_1(r)|$ as the Alfv\'en frequency squared
$\Omega_A^2$ at the distance $r$ and 
estimate that $\Omega_A^2 \ll \Omega_r^2$.
  Evidently the equation for ${\cal E}_1$ corresponds to
free oscillations.  
  It has Lindblad radii approximately
where ${\cal D}(r)\equiv \Omega_r^2-(\Omega_\phi-\omega)^2 =0$.
   For $\nu_*=\omega/\pi =600$ Hz 
and the other parameters the same as
in  Figure 2, the inner Lindblad 
radius is at $r_{Li}/r_S =4.2$
while the outer Lindblad radius is 
at a large distance ($r/r_S>6$). 
  This mode may be driven at $r_{Li}$
by noise or fluctuations in the disk at
this radius, and it is a leading spiral wave.
   The mode amplitude can in principle be found
using three-dimensional MHD
simulations (Koldoba, et al. 2002; Romanova et al. 2007).

   The influence of the magnetically coupled modes
on the flux follows the discussion of \S 2.3.
We find
  $\delta  L \sim \int rdr d\phi
\exp(-i\phi) S(r,\phi,t) \sim  
4 \int rdr T_0^3\epsilon_M |{\cal E}_1/r| \delta
T_1(r)\exp[-i(\omega_1-\Omega_*/2)t]$. 
    For $\nu_* > \sim 380$~Hz, we interpret 
$\omega_1-\Omega_*/2$ as the
lower frequency component of twin QPOs.

\section{Conclusions}

We discuss three  modes of accretion disks 
around rotating magnetized neutron stars which may
explain the frequency separations 
of the twin kilo-Hertz QPOs seen in accreting
X-ray binaries.  
  The existence of  these modes requires 
that there be a maximum
in the angular velocity of the accreting material and
that the fluid is in
stable, nearly circular motion near this maximum rather than moving
rapidly towards the star or out of the disk plane
into funnel flows.   
   It is presently not known if these
conditions occur, but  we are exploring 
 this with 3D magnetohydrodynamic
simulations and will report the results elsewhere. 
  The first mode is a  corotation mode which is
radially trapped in the vicinity of the maximum  of the
disk rotation rate and is unstable.  
   A simple dependence is assumed for the angular
rotation rate of the disk $\Omega_\phi(r)$ which
has a maximum at a radius $r_m $ outside the neutron
star.  
   The unstable mode has a frequency $\omega_1$ somewhat
less than max($\Omega_\phi$) and this can
vary by a significant
factor  depending on the state of the disk (e.g.,
the accretion rate).
  We suggest that this  mode is
associated with the upper 
QPO frequency, $\nu_u=\omega_1/2\pi$.
   The second mode  is a magnetically driven
eccentric ($m=1$)  oscillation of the 
disk excited at the inner
Lindblad radius which is in the vicinity
of the maximum of the disk  rotation.
 The star's magnetic field is assumed to be 
the form discussed by Ruderman (2006).
   The Lindblad radii occur only for relatively
slowly rotating stars, $\nu_*<\sim 380$ Hz. 
  For these stars we suggest that
the lower QPO frequency is $\nu_\ell =\nu_u-\nu_*$.
   The third mode, relevant to more rapidly rotating
stars, is a magnetically coupled
eccentric ($m=1$) and an axisymmetric ($m=0$)
radial disk perturbation.  
   It has an
inner Lindblad radius also in the vicinity
of the maximum of the disk rotation.
   For these stars the lower QPO frequency is
$\nu_\ell=\nu_u-\nu_*/2$.
    A problem remaining for future work is
the determination of the saturation amplitudes of
the different modes.

\section*{Acknowledgements}

  We thank an anonymous referee for valuable
criticism of an earlier version of this work.   
This work was supported in
part by NASA grants NAG5-13220 and
NAG5-13060 and by NSF grant AST-0507760.

\end{document}